\documentclass[letterpaper]{article}
\pdfoutput=1 
\usepackage{xr}
\usepackage[table]{xcolor}

\usepackage{tikz}
\usepackage{pgf}
\usepackage{amsmath,mathtools, booktabs, makecell, tabularx}
\usepackage{float}
\floatstyle{plaintop}
\restylefloat{table}
\usepackage{bbm}
\usepackage{enumerate}
\usepackage[authoryear]{natbib}
\usetikzlibrary{arrows,positioning,automata,calc,fit,shapes.geometric,arrows.meta}
\usepackage{geometry} 
\usepackage{abstract}
\usepackage{multirow}   
\usepackage{color, colortbl}
\usepackage{caption}
\usepackage{subcaption}
\usepackage{longtable}

\usepackage{subfiles}
\usepackage{scalerel,stackengine}
\usepackage{lipsum}

\usepackage[english]{babel}
\usepackage{amsthm}
\usepackage{bbm}
\usepackage{blindtext}
\usepackage{dsfont}
\usepackage{algorithm}
\usepackage{algorithmic}
\usepackage[utf8x]{inputenc}

\usepackage{ulem}

\makeatletter
\def\BState{\State\hskip-\ALG@thistlm}
\makeatother
\usepackage{amssymb}
\usepackage[inline]{enumitem}
\usepackage{url} 

\newcommand{\wh}{\widehat}
\newcommand{\whb}{\widehat{\beta}}
\newcommand{\wt}{\widetilde}
\newcommand{\ddb}{\frac{\partial}{\partial \beta}}

\DeclareFontFamily{OT1}{pzc}{}
\DeclareFontShape{OT1}{pzc}{m}{it}{<-> s * [1.10] pzcmi7t}{}
\DeclareMathAlphabet{\mathpzc}{OT1}{pzc}{m}{it}
\usepackage{geometry}
\usepackage{hyperref}
\hypersetup{colorlinks}

  \hypersetup{
colorlinks =true,
citecolor    = blue,
citebordercolor = violet,
filebordercolor=blue,
linkbordercolor=blue
}

\makeatletter
\newcommand*{\addFileDependency}[1]{
  \typeout{(#1)}
  \@addtofilelist{#1}
  \IfFileExists{#1}{}{\typeout{No file #1.}}
}
\makeatother

\theoremstyle{plain}
\newtheorem{theorem}{Theorem}
\newtheorem{lemma}{Lemma}

\newtheorem{assumption}{Assumption}

\newcommand{\eqn}{\begin{eqnarray}}
\newcommand{\ee}{\end{eqnarray}}
\newcommand{\eqnn}{\begin{eqnarray*}}
\newcommand{\een}{\end{eqnarray*}}
\newcommand{\bal}{\begin{align}}
\newcommand{\eal}{\end{align}}
\newcommand{\be}{\begin{eqnarray}}

\newcommand{\nn}{\nonumber}

\definecolor{LightCyan}{rgb}{0.88,1,1}

\def\spacingset#1{\renewcommand{\baselinestretch}%
{#1}\small\normalsize} \spacingset{1}

\renewcommand{\baselinestretch}{1.66}

\title{Doubly Robust Inference for Hazard Ratio under Informative Censoring with Machine Learning
}
\author{Jiyu Luo\thanks{Herbert Wertheim School of Public Health and Human Longevity Science, University of California, San Diego, La Jolla, CA 92093-0112, USA. E-mail: \href{mailto:jil130@ucsd.edu}{\textsf{jil130@ucsd.edu}}.}~~~and~~Ronghui Xu \thanks{Herbert Wertheim School of Public Health and Human Longevity Science, 
Department of Mathematics and
Halicioglu Data Science Institute, University of California, San Diego, La Jolla, CA 92093-0112, USA. E-mail: \href{mailto:rxu@health.ucsd.edu}{\textsf{rxu@health.ucsd.edu}}.}}

\date{}

\begin{document}
\maketitle
\begin{abstract}
     Randomized clinical trials with time-to-event outcomes have traditionally used the log-rank test followed by the Cox proportional hazards (PH) model to estimate the hazard ratio between the treatment groups. These are valid under the assumption that the right-censoring mechanism is non-informative, i.e.~independent of the time-to-event of interest within each treatment group. More generally, the censoring time might depend on additional covariates, and  inverse probability of censoring weighting (IPCW) can be used to correct for the bias resulting from the informative censoring. IPCW requires a correctly specified censoring time model conditional on the treatment and the covariates. Doubly robust inference in this setting has not been plausible previously due to the non-collapsibility of the Cox model. However, with the recent development of data-adaptive machine learning methods we derive an augmented IPCW (AIPCW) estimator that has the following doubly robust (DR) properties: it is model doubly robust, in that it is consistent and asymptotic normal (CAN), as long as one of the two models, one for the failure time and one for the censoring time, is correctly specified; it is also rate doubly robust, in that it is CAN as long as the product of the estimation error rates under these two models is faster than root-$n$. We investigate the AIPCW estimator using extensive simulation in finite samples.  
\end{abstract}

\section{Introduction}
\label{sec1}

In the analysis of time-to-event data, the Cox proportional hazards (PH) model \citep{C1972} has been widely used to  estimate  the hazard ratio (HR) between two treatment groups in a randomized clinical trial, for example. The validity of the  maximum partial likelihood estimator (MPLE) under the PH model relies on the non-informative censoring assumption \citep{FH1991}; that is, the censoring time random variable is independent of the failure time random variable within each treatment group. In practice, this assumption can  be violated which leads to informative censoring, and 
 the censoring time may well depend on additional covariates. 
 This issue was recently highlighted in \cite{VDV2021}, who aimed to develop procedures to select baseline covariates in order to be adjusted for in the Cox regression model. Such adjustment, however, changes the effect estimand, making it difficult to compare across different adjustment sets. Alternatively, the crude or marginal hazard ratio, as it is often referred to in the medical literature, between the two groups can still be consistently estimated using  inverse probability of censoring weighting (IPCW) under the relaxed censoring assumption that the censoring time and the failure time are independent given the additional covariates.


IPCW  was proposed in \cite{RF2000} to correct for bias resulting from informative censoring 
 of the log-rank test and, prior to that, in \cite{R1993}. Up until then, the main body of literature in both applied and theoretical survival analysis had assumed non-informative censoring, given the predictors in a regression model \citep{FH1991}. 
 A separate line of research where IPCW was called for, was under violation of the PH assumption, where it was recognized that the MPLE gave rise to a population quantity that involved the nuisance censoring distribution \citep{X1996, XO2000}. A series of work has since been done to correct for this bias using IPCW approaches, including \cite{BKG2012, HH2012, NG2017, NG2021}. We note that the terminology `IPCW' was not always mentioned in some of these works, which used the (conditional) survival distribution increments as weights in each risk set; but these are algebraically equivalent to the inverse probability of censoring weights. 
 
 The censoring distribution used in IPCW is often modeled parametrically or semiparametrically, 
 and the resulting IPCW estimator is consistent and asymptotically normal (CAN) if the model is correctly specified. 
\cite{NG2017} proposed a  survival tree approach to estimate the conditional censoring distribution given the covariates, but with no theoretical guarantee for inference. In fact, it is known that the resulting estimator is typically biased \citep{BCH2013}.

Doubly robust (DR) approaches were developed when handling missing data  \citep{R1993, RRZ1995, SRR1999,  RRF2000, RR2001, VR2003, BR2005, T2006}. It is called doubly robust because two working models are involved, one for the outcome of interest, and one for the missing data mechanism,  and 
 the estimator is consistent as long as one of the two working models are correctly specified. 
 When IPW is used to handle the missingness (referred to as coarsening), this usually comes down to augmentation with the coarsened data and the resulting DR estimator is an augmented IPW (AIPW) estimator \citep{T2006}. 

Since right censoring in survival data may be framed as a type of coarsening \citep{T2006}, \cite{RR2005} developed an augmented IPCW (AIPCW) approach for censored survival data. 
 For the PH model, however, this approach is not straightforward to apply to. 
As will be seen later, this is mainly  due to the non-collapsibility of the Cox model \citep{MV2013, TR2012, D2021}.

In this paper, we consider simultaneously the regression parameter and the nuisance baseline hazard function 
under the PH model.  
This naturally gives rise to full data estimating equations that are sums of independent and identically distributed (i.i.d.) martingales. 
The augmentation leads to working models for the failure time and the censoring time given the group indicator and the covariates. 
To specify a conditional failure time model that is compatible with the original (marginal) PH model given the group membership only, data adaptive machine learning (ML) or nonparametric methods are needed. 
With cross-fitting  \citep{CCDDHNR2018}, the resulting AIPCW estimator  has doubly robust properties not only in the classical sense, which is referred to as model doubly robust, but  also rate doubly robust \citep{SRR2019, HBX2021}. Here, rate double robustness refers to an estimator being CAN when 
the product of the estimation error rates under the two working models is faster than root-$n$, while either one of them is allowed to be arbitrarily slow.

The rest of the paper is organized as follows. 
In Section \ref{sec1.1}, we state the model and assumption about censoring. In Section \ref{sec2}, we take a missing data approach by constructing the AIPCW score from the full data score, and  provide a detailed algorithm for the cross-fitted AIPCW estimator. Asymptotic properties of the AIPCW estimator
are described in Section \ref{sec3}. In Section \ref{sec4}, we conduct simulations for the AIPCW estimator using different nuisance estimators, and also compare them with the IPCW estimators. Finally, we conclude with discussion  in Section \ref{sec5}. 
Additional materials are provided in the Appendix.

\subsection{Model and assumption} \label{sec1.1}

Let $T$ and $C$ be the failure time and the censoring time, respectively. 
Denote $X = \min(T,C)$, and $\Delta = I(T \leq C)$. 
Denote also $Y(t) = I(X \ge t)$ the at-risk process, and $N(t) = I(X \le t, \Delta = 1)$ the failure event counting process.  
We consider the two-group survival setting where $A$ is a binary group indicator. For a randomized trial, this can be the treatment groups. 
Let $Z$ be a $p$-dimensional vector of baseline covariates. We assume that the data consist of $n$ independent and identically distributed (i.i.d.) copies of the random vectors $O = (X, \Delta, A, Z)$. 
 \begin{assumption} (informative censoring)
$C \perp T \mid (A,Z)$. \label{assump1}
\end{assumption}

We assume the PH model for the two-group survival:
\be\label{cox}
\lambda(t|A) = \lambda_0(t)\exp(\beta A), 
\ee 
where  $\lambda(t|A)$ denotes the group-specific hazard function of $T$, $\beta$ is the log hazard ratio,  and $\lambda_0(t)$ is the baseline hazard function. 


%

\section{Doubly robust inference} \label{sec2}

In this section following  \cite{T2006} 
we treat right censoring as a coarsened data problem. 
We start with a  set of full data score functions under the PH model, and show that when  IPCW is applied  to this set of full data score functions we obtain the familiar IPCW estimator under the Cox model \citep{BKG2012}. 
We then mimic the approach of  \cite{RR2005} to augment the IPCW score functions and arrive at a doubly robust AIPCW estimator. Finally, for inference purposes we introduce cross-fitting and describe the implementation of the cross-fitted AIPCW estimator.

\subsection{Full data score functions}

The full data vector is $(T, A, Z)$. Following the commonly used NPMLE approach for the semiparametric PH model, the unknown parameters are $\beta$ and  $\Lambda_0(t) = \int_0^{t} \lambda_0(u) du$, the cumulative baseline hazard, 
which is discretized to jumps at the observed event times only \citep{NGAS1992}. 


Following 
\cite{FH1991},
define the full data counting process $N_{T}(t) = I(T \le t)$ and the full data at-risk process $Y_{T}(t) = I(T \ge t)$. 
Let
\be
 M_T(t; \beta, \Lambda_0) = N_T(t) - \int_0^t Y_T(u) e^{\beta A} d\Lambda_0(u).
\ee
Then $ M_T(t; \beta, \Lambda_0) $ is the full data martingale with respect to the full data filtration \\
$\mathcal{F}^f_t = \{N_{T}(u)$, $Y_{T}(u^{+}), A, Z;0\leq u \leq t \}$ under model \eqref{cox}.

We have the following full data score functions for a single copy of the data: 
\begin{align}
& D_1^f(\beta, \Lambda_0, t) =  dM_T(t;\beta, \Lambda_0), \nn  \\
& D_2^f(\beta, \Lambda_0) = \int_0^\tau A dM_T(t;\beta, \Lambda_0). \nn 
\end{align}
where $\tau$ is the maximum follow-up time. Note that $D_1^f(\beta, \Lambda_0, t)$ is a martingale difference that is often used in survival analysis; see for example, \cite{LY2004}. 
For each $t$, the true values of the parameters $\beta$ and $\Lambda_0$ satisfy
\be
E\{D_1^f(\beta, \Lambda_0, t)\} = 0 ~~\text{and}~~ E\{D_2^f(\beta, \Lambda_0)\} = 0. \label{param_def}
\ee

\subsection{IPCW score functions}

In survival analysis, it's common to consider the quantity
\be
    M(t) = N(t) - \int_0^t Y(u) e^{\beta A} d\Lambda_0(u).
\ee
Note that it is not a martingale under informative censoring.
We define $S_c(t|A,Z) = P(C\ge t|a, Z)$  the conditional survival function of $C$, $\wt{\Delta}(t) = I(\min(T,t) < C)$, and denote
\begin{align}
    dM^w(t; \beta, \Lambda_0,  S_c) =& {S_c(t|A,Z)}^{-1} \wt{\Delta}(t)dM_T(t;\beta, \Lambda_0) \nonumber \\
    =& {S_c(t|A,Z)}^{-1} \left\{dN(t) - Y(t) e^{\beta A} d\Lambda_0(t) \right \}.  \label{dM_w}
\end{align}
Note that 
expression \eqref{dM_w} gives  the IPCW score functions:
\begin{align}
&D_1^w(\beta, \Lambda_0, t; S_c) = dM^w(t;\beta, \Lambda_0, S_c), \label{D1 ipcw} \\
&D_2^w(\beta, \Lambda_0; S_c) = \int_0^\tau AdM^w(t;\beta, \Lambda_0, S_c). \label{D2 ipcw}
\end{align}

With $n$ copies of i.i.d.~data, 
this gives the following IPCW weighted estimating equations:
\begin{align}
    & \frac{1}{n} \sum_{i=1}^n D_{1i}^w(\beta, \Lambda, t; S_c) = 0, \nn \\
    & \frac{1}{n} \sum_{i=1}^n D_{2i}^w(\beta, \Lambda; S_c) = 0. \nn
\end{align}
After some algebra, the above estimating equations 
can be combined to give the IPCW partial likelihood score equation  \citep{BKG2012}:
\be \label{beta ee ipcw}
\sum_{i=1}^n \int_0^{\tau} \wh{S}_c(t|A_i,Z_i)^{-1} \left\{ A_i - \frac{\wt{S}^{(1)}(\beta, t; \wh{S}_c)}{\wt{S}^{(0)}(\beta, t; \wh{S}_c)} \right\} dN_i(t) = 0,
\ee
where $\wt{S}^{(l)}(\beta,t; S_c) = \sum_{j=1}^n A_j^l S_c(t|A_j,Z_j)^{-1} Y_j(t) e^{\beta A_j}$ for $l=0, 1$,  and $\wh{S}_c(t|A,Z)$ is some consistent estimator of $S_c(t|A,Z)$.

\subsection{AIPCW score functions}

The consistency of the IPCW estimator  relies critically on ${S}_c(t|A,Z)$ being correctly specified. When it is misspecified,  the IPCW estimator  is biased. 
\cite{RR2005} provides an augmentation approach for an IPCW estimator in survival analysis, so that it has the doubly robust property to be detailed later. 
However, their approach cannot be directly applied because we have not only different weights for different individuals in the data set, but also different weights for each risk set. 
 To this end, it is helpful to  augment the martingale {\it increment} in \eqref{dM_w} as follows.  
 
Denote $N_c(t) = I(X \le t, \Delta =0)$ the counting process for the censoring event, and $\Lambda_c(t|A,Z) = \int_0^t S_c(u|A,Z)^{-1} d\{1 - S_c(u|A,Z)\}$  the cumulative hazard function of $C$ given $A,Z$.  
Then 
$M_c(t; S_c) = N_c(t) - \int_0^t Y(u) d\Lambda_c(u|A,Z)$ is the martingale corresponding to the censoring event counting process with respect to its natural history filtration. 
Also denote $S(t|A,Z) = P(T\ge t|A, Z)$, and $F(t|A,Z) = 1 - S(t|A,Z)$.
Define 
\be
  dM^{aug}(t; \beta, \Lambda_0, S, S_c) 
    &=&  dM^w(t; \beta, \Lambda_0, S_c) + \int_0^t E\{dM_T (t; \beta, \Lambda_0)|A,Z,T\ge u\} \frac{dM_c(u; S_c) }{S_c(u|A,Z)}    \label{dM AIPCW} \\
 &=&   \frac{dN(t) - Y(t) d\Lambda_0(t) e^{\beta A}}{S_c(t|A,Z)} - \, J(t;S,S_c)  \left\{ dS(t|A,Z) + S(t|A,Z) e^{\beta A} d\Lambda_0(t) \right\}, \label{dM aipcw final}
\ee
where  $J(t;S, S_c) = \int_0^t S(u|A,Z)^{-1} S_c(u|A,Z)^{-1} dM_{c} (u;S_c)$. 
The last `=' above used the fact that, for $u\leq t$, 
\begin{align}
 E\{ N_T( t) |A,Z,T\ge u\} = P(T\leq t |A,Z,T\ge u) =  \frac{F(t|A,Z)}{S(u|A,Z)}, \label{expectation1}\\ 
 E\{ Y_T( t) |A,Z,T\ge u\} = P(T\geq t |A,Z,T\ge u) =  \frac{S(t|A,Z)}{S(u|A,Z)}. \label{expectation2}
\end{align}

The above leads to the AIPCW score functions:
\begin{align}
&D_1(\beta, \Lambda_0, t; S, S_c) = dM^{aug}(t; \beta, \Lambda_0, S, S_c),\label{D1 aipcw} \\
&D_2(\beta, \Lambda_0; S, S_c) = \int_0^\tau A \cdot dM^{aug} (t; \beta, \Lambda_0, S, S_c). \label{D2 aipcw}
\end{align}

In Theorem~\ref{thm:dr} below, we will show that \eqref{D1 aipcw} and \eqref{D2 aipcw} are doubly robust score functions. We use superscript $o$ to denote 
the truth; for example, $S^o(t|A,Z)$, $S_c^o(t|A,Z)$ and $\Lambda_c^o(t|A,Z)$ denote the true $S(t|A,Z)$, $S_c(t|A,Z)$ and $\Lambda_c(t|A,Z)$, respectively. Also let $\beta^o$ and $\Lambda_0^o$ denote the true values of the parameters of interest. We assume the following:

\begin{assumption}
$S^o(\tau|a,z) > c$ for $a \in \{0,1\}, z \in \mathcal{Z}$ and some $c > 0$. \label{assump2}
\end{assumption}
\begin{assumption}
$S_c^o(\tau|a,z) > c$ for $a \in \{0,1\}, z \in \mathcal{Z}$ and some $c > 0$. \label{assump3}
\end{assumption}

\begin{theorem}\label{thm:dr}
Under Assumptions \ref{assump1}-\ref{assump3}, if either $S = S^o$ or $S_c = S_c^o$,
\be
E\{D_1(\beta^o,\Lambda_0^o, t;S, S_c)\} = 
E\{D_2(\beta^o,\Lambda_0^o;S, S_c)\} = 0. 
\ee
\end{theorem}
The above theorem states that the scores $ (D_1, D_2) $ identifies the true parameters $ (\beta^o,\Lambda_0^o), $ as long as one of the two survival functions, $S(t|A,Z)$ and $S_c(t|A,Z)$, is true.

Given $n$ i.i.d.~data points, we estimate $\beta^o, \Lambda^o$ by solving
\be
\frac{1}{n} \sum_{i=1}^n D_{1i}(\beta, \Lambda_0, t; S, S_c) = 0, \label{EE1}\\
\frac{1}{n}  \sum_{i=1}^n D_{2i}(\beta, \Lambda_0;S, S_c) = 0. \label{EE2}
\ee
Solving for \eqref{EE1} gives
\be
\wt{\Lambda}_0(\beta,t; S, S_c) = \int_0^t \frac{\frac{1}{n} \sum_{i=1}^n  S_c(u|A_i, Z_i)^{-1} dN_i(u) - J_i(u;S,S_c) dS(u|A_i,Z_i)}{\mathcal{S}^{(0)}(\beta,u;S, S_c)},  \label{lambda aipcw}
\ee
where 
\be
\mathcal{S}^{(l)}(\beta,t; S, S_c) = \frac{1}{n} \sum_{i=1}^n A_i^l e^{\beta A_i} \{S_c(u|A_i,Z_i)^{-1}Y_i(t) + J_i(t;S,S_c)S(t|A_i,Z_i)\} 
\ee
for $l=0, 1$. Further define $\bar{A}(\beta,t; S, S_c) = \mathcal{S}^{(1)}(\beta,t; S, S_c)/\mathcal{S}^{(0)}(\beta,t; S, S_c)$.
After plugging \eqref{lambda aipcw} into \eqref{EE2}, we have: 
\be
    U(\beta; S, S_c) = \frac{1}{n} \sum_{i=1}^n \int_0^{\tau} \{S_c(t|A_i, Z_i)^{-1} dN_i(t) - J_i(u;S,S_c) dS(t|A_i,Z_i) \} \{A_i - \bar{A}(\beta, t; S, S_c) \}=0. \label{beta aipcw}
\ee
It's worth noting that like the partial likelihood score equation, \eqref{beta aipcw} is not a sum of $i.i.d$ terms due to $\bar{A}(\beta, t; S, S_c)$. 
As seen from the derivation leading to \eqref{dM aipcw final}, the augmentation to the weighted martingale increment, which is linear in $N(t)$ and $Y(t)$, is the result of augmentation to the weighted $N(t)$ and $Y(t)$, respectively. 
It is apparent that $S_c(t|A_i, Z_i)^{-1} dN_i(t) - J_i(t;S,S_c) dS(t|A_i,Z_i)$ is the augmented weighted $dN_i(t)$,
and the augmented weighted  $Y_i(t)$'s give rise to the quantities $\mathcal{S}^{(l)}(\cdot)$ and $\bar{A}(\cdot)$, which are the analogies of similar quantities under the usual Cox model. For example, $\bar{A}(\beta, t; S, S_c)$ corresponds to  the empirical mean of the treatment random variable $A$ among subjects who fail at time $t$, which we may denote by $\rho(\beta,t)$. 

The quantity $\rho(\beta,t)$ was implied in \cite{RR2005}, as a nuisance parameter, based on the partial likelihood score function. 
It would, however, not be straightforward to construct compatible models for $\rho(\beta,t)$, which is defined on nested risk sets over time. The set of full data estimating functions we consider here, simultaneously for $\beta$ and $\Lambda_0$, on the other hand, lead naturally to models for $S$ and $S_c$. 

\subsection{Cross-fitted AIPCW estimator}\label{sec:cross-fit}

In practice, both survival functions $S(t|A,Z)$ and $S_c(t|A,Z)$ are unknown and need to be estimated by some estimator $\wh{S}(t|A,Z)$ and $\wh{S}_c(t|A,Z)$. Parametric and semiparametric models, like the Cox model and the accelerated failure time (AFT) model, are often applied since their theoretical properties are well-studies and with little requirement on the computing power. However, these models can be misspecified, especially for $S(t|A, Z)$ due to the non-collapsibility of the Cox model. 
ML or nonparametric methods, like splines \citep{G1992, KST1995a} and random survival forest \citep{IKBL2008}, offer a good alternative. 
ML or nonparametric estimators, however, do not have root-$n$ convergence rate, which makes it difficult to conduct inference. We will show that the asymptotic normality can be established if we also apply cross-fitting, where the entire sample is first split into $k$ folds, and for each fold, we estimate the nuisance functions using only the out-of-fold sample.
Details of the cross-fitted AIPCW estimator $\whb$ are described in Algorithm~\ref{alg:1}. Heuristically, cross-fitting works by inducing independence between the nuisance parameter estimators and the rest of the quantities in the scores, thereby allowing asymptotic normality to be established \citep{SRR2019, HBX2021}. 
\begin{algorithm}[!t]
\caption{$k$-fold Cross-fitted AIPCW estimation of $\beta$}
\label{alg:1}
Input: A sample of $n$ observations that are split into $k$ folds of equal size with index sets $\mathcal{I}_1, \mathcal{I}_2, \ldots, \mathcal{I}_k$.
\begin{algorithmic}
\FOR{each fold indexed by $m$}
\STATE obtain estimated nuisance functions $(\wh{S}^{(-m)}, \wh{S}_c^{(-m)})$ using the out-of-fold sample indexed by $\mathcal{I}_{-m} \coloneqq \{1, \ldots, n \}\setminus \mathcal{I}_m$.
\ENDFOR
\end{algorithmic}
Output: $\whb$, the solution to
\be
\frac{1}{n} \sum_{i=1}^n D_{1i}(\beta, \Lambda_0, t; \wh{S}^{(-m(i))}, \wh{S}_c^{(-m(i))}) = 0,\label{cf EE1}\\
\frac{1}{n}  \sum_{i=1}^n D_{2i}(\beta, \Lambda_0;\wh{S}^{(-m(i))}, \wh{S}_c^{(-m(i))}) = 0 \label{cf EE2},
\ee
where $m(i)$  maps observation $i$ to index of the fold it belongs to.
\end{algorithm}


Quantities involving cross-fitting would be slightly different from quantities without cross-fitting, and will involve the estimated nuisance parameters $(\wh{S}^{(-1)}, \wh{S}_c^{(-1)}), \ldots, (\wh{S}^{(-k)}, \wh{S}_c^{(-k)})$. 
Specifically, solving for \eqref{cf EE1}, we will have
\be
\wt{\Lambda}^{cf}_0(\beta,t; \wh{S} , \wh{S}_c) = \int_0^t \frac{\frac{1}{n} \sum_{i=1}^n  \wh{S}_c^{(-m(i))}(u|A_i, Z_i)^{-1} dN_i(u) - J_i(u;\wh{S}^{(-m(i))},\wh{S}^{(-m(i))}_c) d\wh{S}^{(-m(i))}(u|A_i,Z_i)}{\mathcal{S}^{(0)}(\beta,u;\wh{S} , \wh{S}_c)},  \label{cf lambda aipcw}
\ee
with
\be
\mathcal{S}_{cf}^{(l)}(\beta,t; \wh{S} , \wh{S}_c) = \frac{1}{n} \sum_{i=1}^n A_i^l e^{\beta A_i} \{\wh{S}^{(-m(i))}_c(t|A_i,Z_i)^{-1}Y_i(t) + J_i(t;\wh{S}^{(-m(i))},\wh{S}^{(-m(i))}_c)\wh{S}^{-m(i))}(t|A_i,Z_i)\} 
\ee
for $l=0, 1$. 

Also, $\bar{A}_{cf}(\beta,t; \wh{S} , \wh{S}_c) = \mathcal{S}_{cf}^{(1)}(\beta,t;\wh{S} , \wh{S}_c) / \mathcal{S}_{cf}^{(0)}(\beta,t; \wh{S} , \wh{S}_c)$, and after plugging \eqref{cf lambda aipcw} into \eqref{cf EE2}, we have the final cross-fitted AIPCW estimating equation:
\begin{align}
    &U_{cf}(\beta; \wh{S} , \wh{S}_c) \nn \\
    =&\frac{1}{n} \sum_{i=1}^n \int_0^{\tau} \{\wh{S}^{(-m(i))}_c(t|A_i, Z_i)^{-1} dN_i(t) - J_i(t;\wh{S}^{(-m(i))},\wh{S}^{(-m(i))}_c) d\wh{S}^{(-m(i))}(t|A_i,Z_i) \} \{A_i - \bar{A}(\beta, t; \wh{S} , \wh{S}_c) \}. \label{cf beta aipcw}
\end{align}

We solve the cross-fitted AIPCW estimating equation \eqref{cf beta aipcw} using the Newton-Ralphson algorithm.

\section{Asymptotic Properties} \label{sec3}

We will now describe the asymptotic properties of the proposed cross-fitted AIPCW estimator, when estimated using a random sample of size $n$. We first list a few additional assumptions. 

\begin{assumption}\label{assump4}
There exist $S^*(t|a,z)$ and $S_c^*(t|a,z)$ with $S^*(\tau|a,z) > c$ and $S_c^*(\tau|a,z)>c$  for some $c > 0$, such that
    \begin{align}
    \sup_{t \in [0,\tau], a \in \{0,1\}, z \in \mathcal{Z}} |\wh S(t|a,z) - S^*(t|a,z) | = O_p(a_n), \nn \\
    \sup_{t \in [0,\tau], a \in \{0,1\}, z \in \mathcal{Z}} |\wh S_c(t|a,z) - S_c^*(t|a,z) | = O_p(b_n), \nn
    \end{align}
    for some $a_n = o(1)$ and $b_n = o(1)$.
\end{assumption}

\begin{assumption}
For the limits $S^*$ and $S_c^*$, there exists a neighbourhood $\mathcal{B}$ of $\beta^o$ and functions $\mathpzc{s}^{(l)}(\beta,t; S^*, S_c^*)$ for $l = 0,1$ defined on $\mathcal{B} \times [0,\tau]$ such that $\sup_{t \in [0,\tau], \beta \in \mathcal{B}} |\mathcal{S}^{(l)}(\beta,t; S^*, S_c^*) - \mathpzc{s}^{(l)}(\beta,t; S^*, S_c^*)|$ $= o_p(1)$.\label{assump5}
\end{assumption}

\begin{assumption}
    For $l = 0,1$, $\mathpzc{s}^{(l)}(\beta,t; S^*, S_c^*)$ are continuous functions of $\beta \in \mathcal{B}$, uniformly in $t \in [0,\tau]$ and are bounded on $\mathcal{B} \times [0,\tau]$. $\mathpzc{s}^{(0)}(\beta,t; S^*, S_c^*)$ is bounded away from zero in $\mathcal{B} \times [0,\tau]$.
    For all $\beta \in \mathcal{B}$, $t \in [0,\tau]$:
    \be
    \mathpzc{s}^{(1)}(\beta,t; S^*, S_c^*) = \frac{\partial}{\partial \beta} \mathpzc{s}^{(0)}(\beta,t; S^*, S_c^*) = \frac{\partial ^2}{\partial \beta^2} \mathpzc{s}^{(0)}(\beta,t; S^*, S_c^*) .   
    \ee
    In addition, let $\bar{a} = \mathpzc{s}^{(1)}/ \mathpzc{s}^{(0)}$ and $v = \bar{a} - \bar{a}^2$. We have 
     $\nu(\beta^o; S^*, S_c^*) = \int_0^\tau v(\beta^o, t; S^*, S_c^*)\mathpzc{s}^{(0)}(\beta^o, u; S^*, S_c^*) d\Lambda_0^o(t) > 0.$
    \label{assump6}
\end{assumption}

Assumption \ref{assump4} assumes that both $\wh{S}$ and $\wh{S}_c$ converge to some limiting function $S^*$ and $S_c^*$ that are not necessarily the truth. Here, we do not make the root-$n$ convergence assumption for each of $\wh{S}$ and $\wh{S}_c$ that often limits us to parametric or semiparametric models. This assumption also implies that $\wh{S}^{(-m)}$ and $\wh{S}^{(-m)}_c$  converge to $S^*$ and $S_c^*$ at the same  rate. 
Assumptions \ref{assump5} and \ref{assump6} are similar to regularity assumptions that are typically made under the PH models \citep{AG1982}.

The asymptotic properties of the cross-fitted AIPCW estimator $\whb$ defined in Algorithm~\ref{alg:1} are summarized in  Theorems \ref{thm:consistency} and \ref{thm:AN} below. 
\begin{theorem}\label{thm:consistency}
Under Assumptions \ref{assump4}-\ref{assump6}, if either $S^* = S^o$ or $S_c^* = S_c^o$, then $\whb \overset{p}{\to} \beta^o$. 
\end{theorem}

\begin{theorem}\label{thm:AN}
Under Assumptions $\ref{assump4}$-$\ref{assump6}$, if any of the following conditions hold:

(a) (Rate Double Robustness) $S^* = S^o, S_c^* = S_c^o$ and $a_nb_n = o(n^{-1/2})$;

(b) (Model Double Robustness) $S^* = S^o$ and $a_n = O(n^{-1/2})$. In particular, there exists an influence function $\xi(t,a,z)$ such that $\wh{S}(t|a,z) - S^*(t|a,z) = \sum_{j=1}^n \xi_{j}(t,a,z) /n + o_p(n^{-1/2})$;

(c) (Model Double Robustness) $S_c^* = S_c^o$ and $b_n = O(n^{-1/2})$. In particular, there exists an influence function $\eta(t,a,z)$ such that $\wh{S}_c(t|a,z) - S_c^*(t|a,z) =  \sum_{j=1}^n \eta_{j}(t,a,z) /n + o_p(n^{-1/2})$,

then we have 
\be
\sqrt{n}(\whb - \beta^o) = \frac{1}{\sqrt{n}}\sum_{i=1}^n \nu(\beta^o, S^*, S_c^*)^{-1} \psi_i(\beta^o, \Lambda_0^o,S^*, S_c^*) + o_p(1),  
\ee
where the expression for $\psi_i(\beta^o, \Lambda_0^o,S^*, S_c^*)$ is provided in Appendix~\ref{appendix:notations}.
\end{theorem}

Theorem \ref{thm:AN} establishes both the model double robustness and the rate double robustness properties. Traditionally, doubly robust inference is established assuming both working models are parametric or semiparametric. Model double robustness here allows estimation under the possibly wrong model to converge at any rate. The theorem also establishes rate double robustness, which states that if the estimators under both working models converge to the truth and that their product rate is faster than root-$n$, the proposed AIPCW estimator is CAN even if one of the nuisance estimators converges arbitrarily slowly. This result permits more flexible ML or nonparametric methods with valid inference.

The asymptotic variance of the proposed estimator is simplified under condition (a). In this case, we provide an estimator of the asymptotic variance, which is given in the Theorem~\ref{thm:variance estimator} below. 

\begin{theorem}\label{thm:variance estimator}
Under Assumptions~\ref{assump4}-\ref{assump6}, if condition (a) of Theorem~\ref{thm:AN} holds, i.e. if $S^* = S^o$, $S_c^* = S_c^o$ and $a_nb_n = o_p(n^{-1/2})$, then $\wh{\nu}^{-2}K/n$ is a consistent estimator for the asymptotic variance of $\whb$,
where $\wh{\nu}$ and $K$ are provided in Appendix~\ref{appendix:notations}.
\end{theorem}

When one of the working models is misspecified, the asymptotic variance is rather complicated. 
In this case,  resampling methods such as bootstrap \citep{E1979} may be used to estimate the variance 
since the AIPCW estimator is asymptotically linear.

\section{Simulation}\label{sec4}

In this section, we compare the performance of the cross-fitted AIPCW estimators $\whb$ using different working models,  against different IPCW estimators and the MPLE. 
We consider sample sizes $n=500$ and $n=1000$, and 1000 data sets are  simulated for each setting, which corresponds to margin of error of about $+/- 1.35\%$ for the coverage probability of nominal $95\%$ confidence intervals. Five-fold cross-fitting is used. 

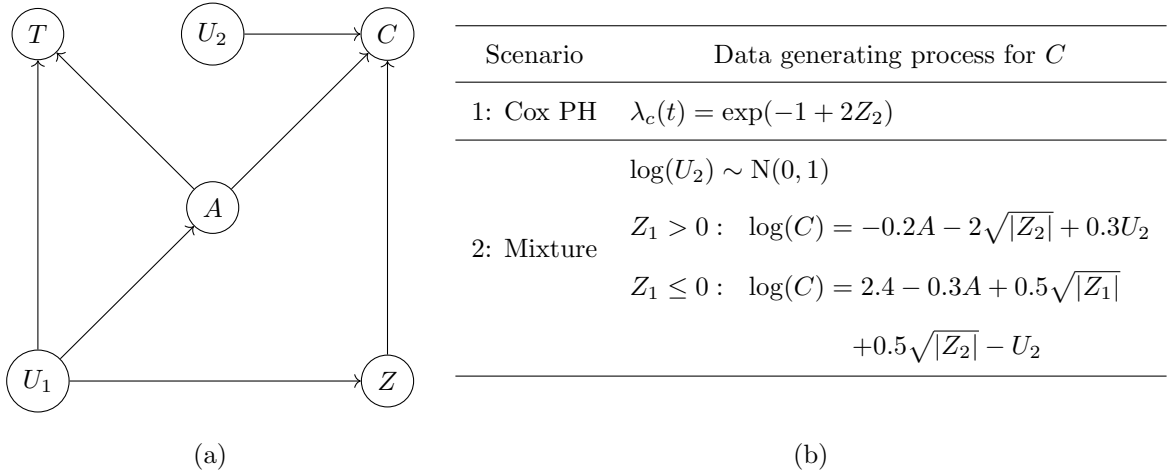
\begin{figure}[ht]
    \centering
    \setcellgapes{5pt}
    \makegapedcells 
    \setlength\belowrulesep{0pt}   
    \setlength\aboverulesep{0pt}
\begin{tabular}{cc}
    \raisebox{-0.5\height}{
    \begin{tikzpicture} \label{dag}
    \node[circle,draw] (v0) at (-2.3cm,2.3cm) {$T$};
    \node[circle,draw] (v1) at (2.3cm, 2.3cm) {$C$};
    \node[circle,draw] (v2) at (-0cm,0cm) {$A$};
    \node[circle,draw] (v3) at (2.3cm,-2.3cm) {$Z$};
    \node[circle,draw] (v4) at (-2.3cm,-2.3cm) {$U_1$};
    \node[circle,draw] (v5) at (0cm,2.3cm) {$U_2$};
    \draw [->] (v2) edge (v1);
    \draw [->] (v3) edge (v1);
    \draw [->] (v4) edge (v0);
    \draw [->] (v4) edge (v3);
    \draw [->] (v4) edge (v2);
    \draw [->] (v2) edge (v0);
    \draw [->] (v5) edge (v1);
    \end{tikzpicture}
    } & \begin{tabular}{ll} \hline
\multicolumn{1}{c}{Scenario}  & \multicolumn{1}{c}{Data generating process for $C$}   \\ \hline
 \multirow{1}{*}{1: Cox PH}   & $\lambda_c(t) = \exp( -1 + 2Z_2)$ \\
 \hline
 \multirow{4}{*}{2: Mixture}   & $\log (U_2) \sim \mbox{N}(0,1)$ \\
 & $Z_1 > 0:~~ \log(C) = - 0.2A - 2\sqrt{|Z_2|} + 0.3U_2$  \\ 
 & $Z_1 \le 0:~~ \log(C) = 2.4 - 0.3A + 0.5\sqrt{|Z_1|}$ \\
 & $~~~~~~~~~~~~~~~~~~~~~~~~~ + 0.5\sqrt{|Z_2|} - U_2$ \\
\hline \end{tabular} \\
(a)  & (b)
\end{tabular}
\caption{(a) Variable diagram. (b) Data generating process for $C$.}
    \label{DGP}
\end{figure}

For  data generation, we first follow the diagram in Figure \ref{DGP}(a) and generate 
 $U_1 \sim$ {Unif} (-1, 1), $A \sim$ {Bernoulli} (0.5), $Z_1 \sim N(0.5U_1, 1)$, $Z_2 \sim  N(U_1^2, 0.09)$,
 and $T = -\log(0.5U_1 + 0.5)e^{-A}$. Here, $T$ follows the PH model  \eqref{cox} with $\beta^o = -1$ and $\lambda_0^o(t) = 1$. 
 
 We consider two scenarios of data generating process for the censoring time $C$, as described in Figure \ref{DGP}(b). 
 Both scenarios  have around $25\%$ samples administratively censored at $\tau = 1$, and $40\%$ of the remaining samples censored during follow-up.  Note that  administrative censoring works in the same way for $T$ and $C$, i.e.~those events are consider as `censored' for both the estimation of $S$ and the estimation of $S_c$. 
It is obvious that Scenario 1 can be correctly modeled.  Scenario 2 is designed such that
most commonly used semiparametric models fail. 
As it turns out, under Scenario 2 
$S_c(\tau |A,Z)$ can be very close to zero for some values of $A$ and $Z$, 
leading to possible violation 
of Assumptions \ref{assump2} and \ref{assump3}.
This echoes 
the argument made in \cite{DDFLS2021} that the overlap assumption needed for DR estimates often fails in practice.

We consider three types of working models: PH model using the R package `survival'; splines \citep{KST1995a} using the R package `polspline'; and random survival forest (RSF) \citep{IKBL2008} using the R package `randomForestSRC'. 
We set splitrule = 'bs.gradient' for RSF, while keeping all the others settings as default. 
We study 7 different combinations of working models for the proposed AIPCW estimator: Cox-Cox, Cox-spline, Cox-RSF, spline-Cox, RSF-Cox, spline-spline, and RSF-RSF, 
where the first part in the names denotes the model for $S$ and the second part denotes the model for $S_c$. 
It is worth noting that due to the non-collapsibility of the Cox model, a semiparametric conditional model for $S$ is almost always misspecified. 
Therefore the consistency of AIPCW-Cox-Cox, AIPCW-Cox-spline and AIPCW-Cox-RSF relies on the correct specification of the censoring model. 
We also note that the convergence rate of the spline and RSF is largely unknown, which  depends on the choice of tuning parameters. 
See Discussion for more on this.

We also investigate the performance of MPLE and various IPCW estimators: IPCW-Cox, IPCW-spline, IPCW-RSF, IPCW-A and IPCW-1. More specifically, IPCW-A  estimates $S_c$ using the product-limit estimator for each group indicated by $A$, while IPCW-1  estimates $S_c$ using the product-limit estimator on the entire sample. 
Robust variance estimator from \cite{BKG2012} is used to estimate the model standard errors of the IPCW estimators. Standard errors for the cross-fitted AIPCW estimators are estimated using Theorem \ref{thm:variance estimator}, which assumes both $S$ and $S_c$ models are correctly specified. 

To avoid numerical problems, 
we impose a minimum on $\wh{S}^{(-m)}(t|A,Z)$ and $\wh{S}^{(-m)}_c(t|A,Z)$ in the above, so that values below 0.01 are trimmed to be 0.01.
Finally, as a benchmark, we also fit model \eqref{cox} to the full data without censoring.

The simulation results for Scenarios 1 and 2 are reported in Tables \ref{sim.1} and \ref{sim.2}, respectively. 
It is immediate that under informative censoring, MPLE, IPCW-1 and IPCW-A have substantial bias leading to poor coverage of the confidence intervals (CI). Under Scenario 1 where the censoring model is correctly specified as Cox, the other three IPCW estimators (-Cox, -spline, -RSF) all appear to perform reasonably well. 
All seven AIPCW estimators also perform well under Scenario 1, with AIPCW-Cox-RSF having larger bias compared to the rest. 

Under Scenario 2,  IPCW-Cox appears more biased than IPCW-spline and IPCW-RSF, as expected. But even for the latter two estimators, their SE's severely under-estimate the SD's, leading to poor coverage of the CI's. 
This also points to the known fact that inference is not guaranteed when ML or nonparametric methods are used in IPCW, as discussed earlier. 
AIPCW-Cox-Cox also has large bias under Scenario 2, as expected. 
The rest six AIPCW's 
are less biased. 
For the larger sample size $n=1000$, AIPCW using two ML or nonparametric methods appears to have the least bias, with close to nominal coverage probabilities. 
Finally we note that, under Scenario 2, spline-based AIPCWs tend to have larger variance. This might be explained by the fact that splines are less stable near the boundary $\tau$, which under Scenario 2 has small $\wh{S}_c(\tau|A,Z)$ for some values of $A$ and $Z$ as mentioned earlier.



\begin{table}[htbp] \begin{center}  
\renewcommand{\arraystretch}{0.93}
\begin{tabular}{llllll} \hline \multicolumn{1}{c}{Sample Size} & \multicolumn{1}{c}{Estimators} & \multicolumn{1}{c}{Bias} & \multicolumn{1}{c}{SD} & \multicolumn{1}{c}{SE} & \multicolumn{1}{c}{CP}\\  \hline  
\multirow{13}{*}{$n=500$}  &  AIPCW-{\color{red} Cox}-Cox & \multicolumn{1}{c}{~0.002~} & \multicolumn{1}{c}{~0.196~} & \multicolumn{1}{c}{~0.191~} & \multicolumn{1}{c}{~0.94~} \\ 
 &  AIPCW-{\color{red} Cox}-spline & \multicolumn{1}{c}{-0.001~} & \multicolumn{1}{c}{~0.198~} & \multicolumn{1}{c}{~0.190~} & \multicolumn{1}{c}{~0.94~} \\
 &  AIPCW-{\color{red} Cox}-RSF & \multicolumn{1}{c}{~0.023~} & \multicolumn{1}{c}{~0.197~} & \multicolumn{1}{c}{~0.207~} & \multicolumn{1}{c}{~0.96~} \\
 &  AIPCW-spline-Cox & \multicolumn{1}{c}{~0.005~} & \multicolumn{1}{c}{~0.185~} & \multicolumn{1}{c}{~0.177~} & \multicolumn{1}{c}{~0.94~} \\
 &  AIPCW-RSF-Cox & \multicolumn{1}{c}{~0.005~} & \multicolumn{1}{c}{~0.189~} & \multicolumn{1}{c}{~0.178~} & \multicolumn{1}{c}{~0.94~} \\
 &  AIPCW-spline-spline & \multicolumn{1}{c}{~0.002~} & \multicolumn{1}{c}{~0.185~} & \multicolumn{1}{c}{~0.177~} & \multicolumn{1}{c}{~0.94~} \\
 &  AIPCW-RSF-RSF & \multicolumn{1}{c}{~0.002~} & \multicolumn{1}{c}{~0.192~} & \multicolumn{1}{c}{~0.190~} & \multicolumn{1}{c}{~0.95~} \\ \cline{2-6}
 &  IPCW-Cox & \multicolumn{1}{c}{-0.006~} & \multicolumn{1}{c}{~0.186~} & \multicolumn{1}{c}{~0.179~} & \multicolumn{1}{c}{~0.94~} \\
 &  IPCW-spline & \multicolumn{1}{c}{-0.005~} & \multicolumn{1}{c}{~0.188~} & \multicolumn{1}{c}{~0.179~} & \multicolumn{1}{c}{~0.94~} \\
 &  IPCW-RSF & \multicolumn{1}{c}{~0.008~} & \multicolumn{1}{c}{~0.190~} & \multicolumn{1}{c}{~0.177~} & \multicolumn{1}{c}{~0.93~} \\
 &  IPCW-{\color{red} A} & \multicolumn{1}{c}{-0.221~} & \multicolumn{1}{c}{~0.180~} & \multicolumn{1}{c}{~0.162~} & \multicolumn{1}{c}{~0.70~} \\
 &  IPCW-{\color{red} 1} & \multicolumn{1}{c}{-0.221~} & \multicolumn{1}{c}{~0.179~} & \multicolumn{1}{c}{~0.162~} & \multicolumn{1}{c}{~0.70~} \\
 &  {\color{red} MPLE} & \multicolumn{1}{c}{-0.205~} & \multicolumn{1}{c}{~0.175~} & \multicolumn{1}{c}{~0.167~} & \multicolumn{1}{c}{~0.76~} \\ \cline{2-6}
  &  Full data & \multicolumn{1}{c}{~0.002} & \multicolumn{1}{c}{~0.103~} & \multicolumn{1}{c}{~0.099~} & \multicolumn{1}{c}{~0.93~} \\ \hline
\multirow{13}{*}{$n=1000$}  &  AIPCW-{\color{red} Cox}-Cox & \multicolumn{1}{c}{-0.008~} & \multicolumn{1}{c}{~0.137~} & \multicolumn{1}{c}{~0.134~} & \multicolumn{1}{c}{~0.94~} \\ 
 &  AIPCW-{\color{red} Cox}-spline & \multicolumn{1}{c}{-0.010~} & \multicolumn{1}{c}{~0.138~} & \multicolumn{1}{c}{~0.133~} & \multicolumn{1}{c}{~0.94~} \\
 &  AIPCW-{\color{red} Cox}-RSF & \multicolumn{1}{c}{~0.019~} & \multicolumn{1}{c}{~0.141~} & \multicolumn{1}{c}{~0.153~} & \multicolumn{1}{c}{~0.97~} \\
 &  AIPCW-spline-Cox & \multicolumn{1}{c}{~0.001~} & \multicolumn{1}{c}{~0.127~} & \multicolumn{1}{c}{~0.123~} & \multicolumn{1}{c}{~0.94~} \\
 &  AIPCW-RSF-Cox & \multicolumn{1}{c}{~0.002~} & \multicolumn{1}{c}{~0.130~} & \multicolumn{1}{c}{~0.125~} & \multicolumn{1}{c}{~0.94~} \\
 &  AIPCW-spline-spline & \multicolumn{1}{c}{~0.001~} & \multicolumn{1}{c}{~0.127~} & \multicolumn{1}{c}{~0.123~} & \multicolumn{1}{c}{~0.94~} \\
 &  AIPCW-RSF-RSF & \multicolumn{1}{c}{-0.005~} & \multicolumn{1}{c}{~0.134~} & \multicolumn{1}{c}{~0.134~} & \multicolumn{1}{c}{~0.95~} \\ \cline{2-6}
 &  IPCW-Cox & \multicolumn{1}{c}{-0.009~} & \multicolumn{1}{c}{~0.130~} & \multicolumn{1}{c}{~0.128~} & \multicolumn{1}{c}{~0.94~} \\
 &  IPCW-spline & \multicolumn{1}{c}{-0.007~} & \multicolumn{1}{c}{~0.135~} & \multicolumn{1}{c}{~0.128~} & \multicolumn{1}{c}{~0.94~} \\
 &  IPCW-RSF & \multicolumn{1}{c}{~0.011~} & \multicolumn{1}{c}{~0.134~} & \multicolumn{1}{c}{~0.128~} & \multicolumn{1}{c}{~0.95~} \\
 &  IPCW-{\color{red} A} & \multicolumn{1}{c}{-0.225~} & \multicolumn{1}{c}{~0.126~} & \multicolumn{1}{c}{~0.114~} & \multicolumn{1}{c}{~0.51~} \\
 &  IPCW-{\color{red} 1} & \multicolumn{1}{c}{-0.224~} & \multicolumn{1}{c}{~0.126~} & \multicolumn{1}{c}{~0.114~} & \multicolumn{1}{c}{~0.51~} \\
 &  {\color{red} MPLE} & \multicolumn{1}{c}{-0.207~} & \multicolumn{1}{c}{~0.122~} & \multicolumn{1}{c}{~0.118~} & \multicolumn{1}{c}{~0.58~} \\ \cline{2-6}
 &  Full data & \multicolumn{1}{c}{-0.003~} & \multicolumn{1}{c}{~0.069~} & \multicolumn{1}{c}{~0.07~} & \multicolumn{1}{c}{~0.94~} \\
\hline \end{tabular} \end{center} 
{\footnotesize SD: standard deviation; SE: standard error; CP: coverage probability of nominal 95\% CI}
\captionsetup{width=.9\linewidth}  \caption{Simulation results under Scenario 1. Data are generated following Figures \ref{DGP}(a) and (b) with $\beta^o = -1$. {\color{red} Red} indicates that the model or approach is invalid. 
} \label{sim.1} 
\end{table}

\begin{table}[htbp] \begin{center}  
\renewcommand{\arraystretch}{0.93}
\begin{tabular}{llllll} \hline \multicolumn{1}{c}{Sample Size} & \multicolumn{1}{c}{Estimators} & \multicolumn{1}{c}{Bias} & \multicolumn{1}{c}{SD} & \multicolumn{1}{c}{SE} & \multicolumn{1}{c}{Coverage}\\  \hline  
\multirow{13}{*}{$n=500$}  &  AIPCW-{\color{red} Cox}-{\color{red} Cox} & \multicolumn{1}{c}{-0.129~} & \multicolumn{1}{c}{~0.285~} & \multicolumn{1}{c}{~0.276~} & \multicolumn{1}{c}{~0.93~} \\ 
 &  AIPCW-{\color{red} Cox}-spline & \multicolumn{1}{c}{-0.029~} & \multicolumn{1}{c}{~0.604~} & \multicolumn{1}{c}{~0.623~} & \multicolumn{1}{c}{~0.97~} \\
 &  AIPCW-{\color{red} Cox}-RSF & \multicolumn{1}{c}{-0.064~} & \multicolumn{1}{c}{~0.249~} & \multicolumn{1}{c}{~0.243~} & \multicolumn{1}{c}{~0.93~} \\
 &  AIPCW-spline-{\color{red} Cox} & \multicolumn{1}{c}{-0.068~} & \multicolumn{1}{c}{~0.282~} & \multicolumn{1}{c}{~0.256~} & \multicolumn{1}{c}{~0.93~} \\
 &  AIPCW-RSF-{\color{red} Cox} & \multicolumn{1}{c}{-0.034~} & \multicolumn{1}{c}{~0.275~} & \multicolumn{1}{c}{~0.250~} & \multicolumn{1}{c}{~0.93~} \\
 &  AIPCW-spline-spline & \multicolumn{1}{c}{~0.038~} & \multicolumn{1}{c}{~0.578~} & \multicolumn{1}{c}{~0.585~} & \multicolumn{1}{c}{~0.96~} \\
 &  AIPCW-RSF-RSF & \multicolumn{1}{c}{-0.039~} & \multicolumn{1}{c}{~0.264~} & \multicolumn{1}{c}{~0.238~} & \multicolumn{1}{c}{~0.93~} \\ \cline{2-6}
 &  IPCW-{\color{red} Cox} & \multicolumn{1}{c}{-0.114~} & \multicolumn{1}{c}{~0.266~} & \multicolumn{1}{c}{~0.174~} & \multicolumn{1}{c}{~0.77~} \\
 &  IPCW-spline & \multicolumn{1}{c}{-0.046~} & \multicolumn{1}{c}{~0.452~} & \multicolumn{1}{c}{~0.192~} & \multicolumn{1}{c}{~0.68~} \\
 &  IPCW-RSF & \multicolumn{1}{c}{-0.088~} & \multicolumn{1}{c}{~0.257~} & \multicolumn{1}{c}{~0.179~} & \multicolumn{1}{c}{~0.80~} \\
 &  IPCW-{\color{red} A} & \multicolumn{1}{c}{-0.227~} & \multicolumn{1}{c}{~0.184~} & \multicolumn{1}{c}{~0.170~} & \multicolumn{1}{c}{~0.74~} \\
 &  IPCW-{\color{red} 1} & \multicolumn{1}{c}{-0.226~} & \multicolumn{1}{c}{~0.183~} & \multicolumn{1}{c}{~0.166~} & \multicolumn{1}{c}{~0.72~} \\
 &  {\color{red} MPLE} & \multicolumn{1}{c}{-0.216~} & \multicolumn{1}{c}{~0.179~} & \multicolumn{1}{c}{~0.174~} & \multicolumn{1}{c}{~0.77~} \\ \cline{2-6}
 &  Full data & \multicolumn{1}{c}{~0.002} & \multicolumn{1}{c}{~0.103~} & \multicolumn{1}{c}{~0.099~} & \multicolumn{1}{c}{~0.93~} \\
 \hline
\multirow{13}{*}{$n=1000$}  &  AIPCW-{\color{red} Cox}-{\color{red} Cox} & \multicolumn{1}{c}{-0.127~} & \multicolumn{1}{c}{~0.195~} & \multicolumn{1}{c}{~0.192~} & \multicolumn{1}{c}{~0.90~} \\ 
 &  AIPCW-{\color{red} Cox}-spline  & \multicolumn{1}{c}{-0.056~} & \multicolumn{1}{c}{~0.396~} & \multicolumn{1}{c}{~0.367~} & \multicolumn{1}{c}{~0.95~} \\
 &  AIPCW-{\color{red} Cox}-RSF  & \multicolumn{1}{c}{-0.035~} & \multicolumn{1}{c}{~0.187~} & \multicolumn{1}{c}{~0.189~} & \multicolumn{1}{c}{~0.95~} \\
 &  AIPCW-spline-{\color{red} Cox}  & \multicolumn{1}{c}{-0.056~} & \multicolumn{1}{c}{~0.191~} & \multicolumn{1}{c}{~0.180~} & \multicolumn{1}{c}{~0.93~} \\
 &  AIPCW-RSF-{\color{red} Cox}  & \multicolumn{1}{c}{-0.021~} & \multicolumn{1}{c}{~0.185~} & \multicolumn{1}{c}{~0.178~} & \multicolumn{1}{c}{~0.92~} \\
 &  AIPCW-spline-spline   & \multicolumn{1}{c}{~0.008~} & \multicolumn{1}{c}{~0.344~} & \multicolumn{1}{c}{~0.332~} & \multicolumn{1}{c}{~0.95~} \\
 &  AIPCW-RSF-RSF   & \multicolumn{1}{c}{-0.020~} & \multicolumn{1}{c}{~0.198~} & \multicolumn{1}{c}{~0.179~} & \multicolumn{1}{c}{~0.93~} \\ \cline{2-6}
 &  IPCW-{\color{red} Cox} & \multicolumn{1}{c}{-0.103~} & \multicolumn{1}{c}{~0.204~} & \multicolumn{1}{c}{~0.126~} & \multicolumn{1}{c}{~0.71~} \\
 &  IPCW-spline  & \multicolumn{1}{c}{-0.045~} & \multicolumn{1}{c}{~0.377~} & \multicolumn{1}{c}{~0.146~} & \multicolumn{1}{c}{~0.63~} \\
 &  IPCW-RSF  & \multicolumn{1}{c}{-0.047~} & \multicolumn{1}{c}{~0.202~} & \multicolumn{1}{c}{~0.134~} & \multicolumn{1}{c}{~0.78~} \\
 &  IPCW-{\color{red} A} & \multicolumn{1}{c}{-0.220~} & \multicolumn{1}{c}{~0.127~} & \multicolumn{1}{c}{~0.120~} & \multicolumn{1}{c}{~0.56~} \\
 &  IPCW-{\color{red} 1} & \multicolumn{1}{c}{-0.219~} & \multicolumn{1}{c}{~0.127~} & \multicolumn{1}{c}{~0.117~} & \multicolumn{1}{c}{~0.53~} \\
 &  {\color{red} MPLE} & \multicolumn{1}{c}{-0.211~} & \multicolumn{1}{c}{~0.123~} & \multicolumn{1}{c}{~0.123~} & \multicolumn{1}{c}{~0.61~} \\ \cline{2-6}
 &  Full data & \multicolumn{1}{c}{-0.003~} & \multicolumn{1}{c}{~0.069~} & \multicolumn{1}{c}{~0.07~} & \multicolumn{1}{c}{~0.94~} \\
\hline \end{tabular} \end{center} 
{\footnotesize SD: standard deviation; SE: standard error; CP: coverage probability of nominal 95\% CI}
\captionsetup{width=.9\linewidth}  \caption{Simulation results under Scenario 2. Data are generated following Figures \ref{DGP}(a) and (b) with $\beta^o = -1$. {\color{red} Red} indicates that the model or approach is invalid. 
} \label{sim.2} 
\end{table}

\newpage
\section{Discussion} \label{sec5}

For the analysis of two-group survival, including for 
randomized clinical trials, non-informative censoring is assumed. When the simple PH model \eqref{cox} is used with no covariates adjusted for, this requires the censoring distribution to be independent of any covariates. When this assumption is 
violated, 
the commonly used MPLE is biased and typically IPCW is used to correct that bias if the interest remains to estimate the marginal hazard ratio between the two groups. IPCW, on the other hand, requires  modeling the censoring distribution, which can be wrong unless ML or nonparametric estimates are used. 
In this paper we have developed an 
AIPCW estimator that is both model DR and rate DR. Rate double robustness allows us to get around the non-collapsibility of the Cox regression model using more flexible ML or nonparametric methods for the conditional failure time model  demanded by the DR construct,  because almost any parametric or semiparametric  would otherwise be invalid.

The theoretical results require certain rate condition of the estimates of the nuisance parameters. These are not always established for a given ML or nonparametric estimator. 
\cite{CZZK2022} and \cite{KST1995b} demonstrated that under certain conditions, rate better than $n^{1/4}$ can be achieved for random survival forest and splines, which would lead to a faster than root-$n$ product rate. 
Faster than $n^{1/4}$ rate are also shown to be attainable for other ML methods, for example, regression trees \citep{WW2015} and neural networks \citep{CW1999}. 
The rates, of course, depend on the hyper-parameter values.
In the simulations we used the default settings for the spline and the random survival forest. Investigation of other ML or nonparametric methods, as well as their tuning, in relationship with the performance of DR estimators, remains a topic of future work. 

This work focused on two-group survival and a  binary  $A$.
Generalization to continuous and/or multivariate $A$ is conceptually straightforward although different algebra might be involved.
In particular for continuous $A$, 
we would no longer have $A^2 =A$ and additional quantities like $\mathcal{S}^{(2)}$ need to be introduced.  

Finally the models for $S$ and $S_c$ may include additional and different  sets of covariates for these two models, 
so long as the failure time and the censoring time are independent given the common covariates $Z$.

The R codes for the cross-fitted AIPCW estimator as well as the simulation procedures investigated in this work are available online in \url{http://github.com/charlesluo1002/DR-Cox}.

\newpage
{\textbf{\Large Appendix} 

\appendix

\section{Notation and Expressions}  \label{appendix:notations}

First, we list or repeat notations that will be used in the proofs. 
For $i$ in $1, \ldots, n$, we define 
\begin{align*}
    & M_{ci}(t; S_c) = N_c(t) - \int_0^t Y(u) d\Lambda_c(u|A_i,Z_i), \\
    & J_i(t;S, S_c) = \int_0^t S(u|A_i,Z_i)^{-1} S_c(u|A_i,Z_i)^{-1} dM_{c} (u;S_c), \\
    & d\mathcal{N}_{i}(t;S, S_c) = S_c(t|A_i, Z_i)^{-1} dN_i(t) - J_i(t;S,S_c) dS(t|A_i,Z_i),\\
    & \Gamma_i^{(l)}(\beta,t; S, S_c) = A_i^l e^{\beta A_i} \{S_c(t|A_i,Z_i)^{-1}Y_i(t) + J_i(t;S,S_c)S(t|A_i,Z_i)\} ,\\
    & \mathcal{S}^{(l)}(\beta,t;S,S_c) = \frac{1}{n} \sum_{i=1}^n \Gamma_i^{(l)}(\beta,t;S,S_c), \\
    & dM^{aug}_i(t;\beta,\Lambda_0,S, S_c) = d\mathcal{N}_{i}(t, S, S_c) - \Gamma_i^{(0)}(\beta,t;S, S_c) d\Lambda_0(t) ,\\
    & \bar{A}(\beta,t; S, S_c) = \mathcal{S}^{(1)}(\beta,t; S, S_c)/\mathcal{S}^{(0)}(\beta,t; S, S_c), \\ 
    & U(\beta; S, S_c) = \frac{1}{n}\sum_{i=1}^n \int_0^{\tau} d\mathcal{N}_{i}(t;S, S_c) \{A_i - \bar{A}(\beta, t; S, S_c) \}, \\
    & V(\beta,t; S, S_c) = d\bar{A}(\beta,t; S, S_c)/d\beta = \bar{A}(\beta,t; S, S_c) - \bar{A}(\beta,t; S, S_c)^2 ,\\
    & \bar{a}(\beta, t; S, S_c) = \mathpzc{s}^{(1)}(\beta, t; S, S_c)/ \mathpzc{s}^{(0)}(\beta, t; S, S_c) ,\\
    & v(\beta, t; S, S_c) = \bar{a}(\beta, t; S, S_c) - \bar{a}(\beta, t; S, S_c)^2, \\
    &\mu(\beta; S, S_c) = \int_0^\tau \{\bar{a}(\beta^o, t; S, S_c) - \bar{a}(\beta, t; S, S_c) \} \mathpzc{s}^{(0)}(\beta^o, t; S, S_c) d\Lambda_0^o(t), \\
    &\nu(\beta; S, S_c) = \int_0^\tau v(\beta, t; S, S_c)\mathpzc{s}^{(0)}(\beta^o, t; S, S_c) d\Lambda_0^o(t).
\end{align*}


\newpage
Next are expressions used in the asymptotic results. 
We stated in Theorem~\ref{thm:AN} that 
\be
\sqrt{n}(\whb - \beta^o) = \frac{1}{\sqrt{n}}\sum_{i=1}^n \nu(\beta^o, S^*, S_c^*)^{-1} \psi_i(\beta^o, \Lambda_0^o,S^*, S_c^*) + o_p(1).  
\ee
Here  $\psi_i (\beta^o, \Lambda_0^o,S^*, S_c^*) 
= \psi_{1i} + \psi_{2i} + \psi_{3i}$ where
\begin{align}
    \psi_{1i}   =&  \int_0^\tau \{A_i - \bar{a}(\beta^o,t;S^*, S_c^*)\}dM^{aug}_i(t;\beta^o,\Lambda_0^o,S^*, S_c^*),  \label{psi1} \\
 \psi_{2i}  =&   \frac{k}{n(k-1)} \sum_{j \in \mathcal{I}_{-m(i)}} \int_0^{\tau} \{\bar{a}(\beta^o,t;S^*, S_c^*) - A_i\}J_i(t;S^*, S_c^*) \{d\xi_{j}(t,A_i, Z_i) +e^{\beta^oA_i} \xi_{j}(t,A_i, Z_i)d\Lambda_0^o(t) \}, \nn \\
\psi_{3i}  =&   \frac{k}{n(k-1)} \sum_{j \in \mathcal{I}_{-m(i)}}  \int_0^{\tau} \{ A_i - \bar{a}(\beta^o,t;S^*, S_c^*)\} \Bigg( \frac{\eta_{j}(t,A_i, Z_i)}{S_c^*(t|A_i,Z_i)^2}\{dN_i(t) - Y_i(t)e^{\beta^oA_i}d\Lambda_0^o(t) \} \nn \\
    &- \int_0^t \bigg [\frac{S^*(u|A_i,Z_i)}{S^*_c(u|A_i,Z_i)^2} \eta_{j}(u,A_i,Z_i) \{dM_{ci}(u;S_c^*) + Y_i(u) S_c^*(u|A_i,Z_i)^{-1} dS_c^*(u|A_i,Z_i) \} \nn\\
    &+ S^*(u|A_i,Z_i)^{-1}S_c^*(u|A_i,Z_i)^{-2}Y_i(u)d\eta_j(u,A_i,Z_i) \bigg ]\{dS^*(t|A_i,Z_i) + S^*(t|A_i,Z_i)e^{\beta^oA_i} d\Lambda_0^o(t) \} \Bigg). \nn
\end{align}
The long expression above simplifies depending on which models are correctly specified. Under case (a) of Theorem~\ref{thm:AN}, when both $S$ and $S_c$ are correctly specified, $ \psi_{2i} = \psi_{3i} =0$. Under case (b), when $\wh{S}$ is $\sqrt{n}$-consistent , $\psi_{3i} = 0$. Under case (c), when $\wh{S}_c$ is $\sqrt{n}$-consistent, $\psi_{2i} = 0$. 

In Theorem~\ref{thm:variance estimator}, we stated that $\wh{\nu}^{-2}K/n$ is a consistent estimator for the asymptotic variance of $\whb$ when both $S^* = S^o$ and $S_c^* = S_c^o$ are correctly specified. 
The expression for $K$ and $\wh{\nu}$ are as follows:
\begin{align*}
    \wh{\nu}=& \frac{1}{n} \sum_{i=1}^n \int_0^\tau V(\whb,t; \wh{S}, \wh{S}_c) d\mathcal{N}_i(t; \wh{S}^{(-m(i))}, \wh{S}^{(-m(i))}_c), \\
    K =& \frac{1}{n} \sum_{i=1}^n \wt{\psi}_{1i}(\whb, \wt{\Lambda}_0(\whb, \cdot;\wh{S}, \wh{S}_c), \wh{S}^{(-m(i))}, \wh{S}^{(-m(i))}_c)^2, 
\end{align*}
where
\be
    \wt{\psi}_{1i}(\beta, \Lambda_0, S, S_c) = \int_0^\tau \{A_i - \bar{A}(\beta,t;S, S_c)\}dM^{aug}_i(t;\beta, \Lambda_0, S, S_c). \nn
\ee

\newpage
\section{Proof of Double Robustness}

\begin{lemma}\label{lem:tsiatis}
For any $S_c(t|A,Z)$ with its corresponding censoring specific martingale $M_c(t;S_c)$,
\be
\int_0^t \frac{dM_c(u;S_c)}{ S_c(u|A,Z)}  = 1 - 
\frac{Y(t)}{ S_c (t|A,Z)} - 
\frac{N(t-)}{ S_c(X|A,Z)}, \label{l1.3}
\ee
where $N(t-) = I(X<t, T\leq C)$.
\end{lemma}
Note, this can be seen as a continuous version of Lemma 10.4 in \cite{T2006}.
\subsubsection*{Proof} 

First note that 
\be
\int_0^t \frac{ dN_c(u) }{S_c(u|A,Z)}
= 
\frac{N_c(t-)}{S_c(X|A,Z)},  
\label{l1.1}  
\ee 
where $N_C(t-) = I(X<t, T > C)$.
Next, since $ S_c(u|A,Z) = \exp  \{-\Lambda_c(u|A,Z) \} $, 
\begin{align}
    &~~~~\int_0^t  \frac {- Y(u) d\Lambda_c(u|A,Z) }{S_c(u|A,Z) } \nonumber \\
    &= I(X \ge t)\int_0^t \frac{dS_c(u|A,Z) }{S_c(u|A,Z)^2} +I(X
    < t)\int_0^X \frac{dS_c(u|A,Z) }{S_c(u|A,Z)^2}  \nonumber\\
    &= I(X \ge t)\{ - S_c(u|A,Z)^{-1} \}|_{u=0}^{u=t} + I(X < t)\{ - S_c(u|A,Z)^{-1} \}|_{u=0}^{u=X}  \nonumber\\
    &= 1 - \frac{ Y(t)}{ S_c(t|A,Z)}  - \frac{ I(X < t) }{S_c(X|A,Z)}. \label{l1.2}
\end{align}
Since $I(X < t) = N(t-) + N_c(t-) $,
\eqref{l1.1} + \eqref{l1.2} then gives the lemma.
\qed

\subsubsection*{Proof of Theorem~\ref{thm:dr}}

Recall that
\be
dM^{aug}(t; \beta, \Lambda_0, S, S_c) = dM^w(t; \beta,\Lambda_0, S_c) - J(t;S,S_c) 
\left\{ dS(t|A,Z) + S(t|A,Z)e^{\beta A}d\Lambda_0(t) \right\}, \nn
\ee
where $J(t;S, S_c)$ is also included in Appendix~\ref{appendix:notations}.

a) Assume $S_c = S_c^o$.

We first consider $dM^w(t;\beta^o,\Lambda_0^o, S_c^o)$. For $h(A) = 1$ or $A$,
\eqnn
  &&  E \left \{ h(A)dM^w(t;\beta^o,\Lambda_0^o,S_c^o)  \right \}  \\
   &=&  E\left\{ h(A) S_c^o(t|A,Z)^{-1} \left[d  E \{ I(T\leq t)I(C\ge t)|T=t,A,Z  \} \right.\right. \\
  && \left.\left. - E\{  I(T \ge t) I(C \ge t)| T=t, A,Z\} \cdot e^{\beta^oA}d\Lambda_0^o(t)  \right] \right\} \\
       &=& E[ h(A) S_c^o(t|A,Z)^{-1} \{ dI(T\leq t)P(C\ge t| A,Z)   \\
    &&-  I(T \ge t) P(C \ge t| A,Z) \cdot e^{\beta^oA}d\Lambda_0^o(t)     \}] \\
    &=&  E\{h(A) dM_T(t;\beta^o, \Lambda_0^o)\} \\
    &=& 0,
\een
where the second `=' above uses the informative censoring Assumption 1. 

Next we consider $J(t;S,S_c) \{ dS(t|A,Z) + S(t|A,Z)e^{\beta^o A}d\Lambda_0^o(t)\}$. Its expectation being zero follows immediately  from the fact that $M_c(t;S_c^o)$ is a martingale. 


b) Assume $S = S^o$. 

Noting that $Y_T(t)N(t-) = N(t-) dN_T(t)=0$ and $Y(t) dN_T(t)= dN(t) $, 
we multiply \eqref{l1.3} by \\
$dM_T(t) = dN_T(t) - Y_T(t)e^{\beta^o A}d\Lambda_0(t)$ giving:
\eqnn
&& dM_T(t;\beta^o, \Lambda_0^o)\int_0^t \frac{dM_c(u;S_c)}{S_c(u|A,Z)} \\
&=&     dN_T(t)\int_0^t \frac{dM_c(u;S_c)}{S_c(u|A,Z)}
- Y_T(t)e^{\beta^o A}d\Lambda_0^o(t)\int_0^t \frac{dM_c(u;S_c)}{S_c(u|A,Z)} \\
&=& dN_T(t) - 
\frac{dN_T(t) Y(t)}{S_c(t|A,Z)} - 
\frac{dN_T(t) N(t-)}{S_c(X|A,Z)} 
- Y_T(t)e^{\beta^o A}d\Lambda_0^o(t) + \frac{Y(t)e^{\beta^o A}d\Lambda_0^o(t) }{S_c(t|A,Z)} + \frac{ Y_T(t)N(t-)e^{\beta^o A}d\Lambda_0^o(t) }{S_c(X|A,Z)}. \\
&=& dM_T(t) - dM^w(t).
\een
Therefore
\be
 dM^w(t;\beta^o, \Lambda_0^o)= dM_T(t;\beta^o, \Lambda_0^o) - 
 dM_T(t;\beta^o, \Lambda_0^o)\int_0^t \frac{dM_c(u;S_c)}{S_c(u|A,Z)}. \nn
\ee

We note that \eqref{expectation1} and \eqref{expectation2} hold when  $S = S^o$. 
From  \eqref{dM AIPCW} then we have
\begin{align*}
&E\{dM^{aug}(t;\beta^o, \Lambda_0^o, S^o,S_c)\} \\
=& E\left[ dM^w(t; \beta^o, \Lambda_0^o, S_c) 
+ \int_0^t E\{dM_T (t; \beta^o, \Lambda_0^o)|A,Z,T\ge u\} \frac{dM_c(u; S_c) }{S_c(u|A,Z)} \right]\\
=& 
E \left[dM_T (t;\beta^o,\Lambda_0^o) - dM_T(t;\beta^o,\Lambda_0^o)\int_0^t \frac{dM_c(u;S_c)}{S_c(u|A,Z)} 
 + \int_0^t E\{dM_T (t; \beta^o, \Lambda_0^o)|A,Z,T\ge u\} \frac{dM_c(u; S_c) }{S_c(u|A,Z)} \right]
\\
 =& 
E \left\{\int_0^t \Big[E\{dM_T (t; \beta^o, \Lambda_0^o)|A,Z,T\ge u\} - dM_T(t;\beta^o,\Lambda_0^o) \Big] \frac{dM_c(u; S_c) }{S_c(u|A,Z)} \right\} \\
  =& 
 E \left[ E\left\{\int_0^t \Big[E\{dM_T (t; \beta^o, \Lambda_0^o)|A,Z,T\ge u\} - dM_T(t;\beta^o,\Lambda_0^o) \Big] \frac{dN_c(u) }{S_c(u|A,Z)} \Big|A,Z,T\ge u, C= u \right\}\right] \\
 &- E \left[ E \left\{\int_0^t \Big[E\{dM_T (t; \beta^o, \Lambda_0^o)|A,Z,T\ge u\} - dM_T(t;\beta^o,\Lambda_0^o) \Big] \frac{Y(u)d\Lambda_c(u)}{S_c(u|A,Z)} \Big|A,Z,T\ge u, C\ge u  \right\}\right] \\
   =& 
E\left\{\int_0^t \frac{dN_c(u) }{S_c(u|A,Z)} \Big[E\{dM_T (t; \beta^o, \Lambda_0^o)|A,Z,T\ge u, C=u\} - E\{dM_T (t; \beta^o, \Lambda_0^o)|A,Z,T\ge u, C=u\} \Big] \right\} \\
 &- E\left\{\int_0^t \frac{Y(u)d\Lambda_c(u)}{S_c(u|A,Z)} \Big[E\{dM_T (t; \beta^o, \Lambda_0^o)|A,Z,T\ge u, C \ge u\} - E\{dM_T (t; \beta^o, \Lambda_0^o)|A,Z,T\ge u, C \ge u\} \Big] \right\} \\
 =& 0,
\end{align*}
where in the 3rd line above $E \{ d M_T (t;\beta^o,\Lambda_0^o) \} = 0$ because $ M_T (t;\beta^o,\Lambda_0^o) $ is a martingale. 

The above also gives 
\be
    E\left\{\int_0^t AdM^{aug}(t;\beta^o, \Lambda_0^o, S^o,S_c) \right\} = 0. \nn
\ee
\qed

\end{document}